\documentclass[9pt,english]{IEEEtran}
\usepackage[T1]{fontenc}
\usepackage[latin9]{inputenc}
\usepackage{float}
\usepackage{mathrsfs}
\usepackage{mathtools}
\usepackage{bm}
\usepackage{bbm}
\usepackage{amsmath}
\usepackage{amsthm}
\usepackage{amssymb}

\usepackage{filecontents} 
\theoremstyle{definition}
\newtheorem{definition}{Definition}

\usepackage{graphicx}
\usepackage{epstopdf}

\usepackage{tabularx,booktabs}
\newcolumntype{C}[1]{>{\centering\arraybackslash}p{#1}}
\usepackage{array,multirow}

\makeatletter

\usepackage{todonotes}

\floatstyle{ruled}
\newfloat{algorithm}{tbp}{loa}
\providecommand{\algorithmname}{Algorithm}
\floatname{algorithm}{\protect\algorithmname}

\theoremstyle{plain}

\theoremstyle{plain}

\theoremstyle{plain}

\usepackage{spconf}
\usepackage[numbers,sort&compress]{natbib}

\@ifundefined{showcaptionsetup}{}{%
 \PassOptionsToPackage{caption=false}{subfig}}
\usepackage{subfig}
\makeatother

\usepackage{babel}
\providecommand{\lemmaname}{Lemma}
\providecommand{\propositionname}{Proposition}
\providecommand{\theoremname}{Theorem}

\begin{document}
\def\thesection{\arabic{section}}

\def\thesubsection{\arabic{section}.\arabic{subsection}}

\def\thesubsubsection{\arabic{section}.\arabic{subsection}.\arabic{subsubsection}}

\makeatletter
\renewcommand\footnoterule{%
  \kern-4\p@
  \hrule\@width.4\columnwidth
  \kern2.6\p@}
  \makeatother

\title{Dictionary Learning Using Rank-One Projection (ROP)}
\maketitle

\begin{abstract}
Dictionary learning aims to find a dictionary that can sparsely represent the training data. Methods in the literature typically formulate the dictionary learning problem as an optimisation with respect to two variables, i.e., dictionary and sparse coefficients, and solve it by alternating between two stages: sparse coding and dictionary update. The key contribution of this work is a Rank-One Projection (ROP) formulation where dictionary learning is cast as an optimisation with respect to a single variable which is a set of rank one matrices. The resulting algorithm is hence single staged. An alternating direction method of multipliers (ADMM) is derived to solve the optimisation problem and guarantees a global convergence despite non-convexity of the optimisation formulation. Also ROP reduces the number of tuning parameters required in other benchmark algorithms. Numerical tests demonstrate that ROP outperforms other benchmarks for both synthetic and real data especially when the sample number is small. 
\end{abstract}


\section{Introduction}

\noindent
Sparse signal representation has drawn extensive attention due to its various applications in signal denoising \cite{elad2006image,dabov2007image}, restoration \cite{mairal2008sparse,dong2013nonlocally},
source separation \cite{li2006underdetermined,abolghasemi2012blind},
classification \cite{tosic2011dictionary,huang2007sparse}, recognition
\cite{wright2009robust,wright2010sparse,zhang2011sparse}, image super-resolution
\cite{yang2010image,dong2011image} to name a few. The basic idea
of sparse signal representation is that an observed signal can be
approximated as a linear combination of a few number of codewords
picking from a dictionary. Compared with choosing a basis set from predefined dictionaries such as Fourier and wavelet transforms, a dictionary trained from the data itself can attain sparser representations \cite{olshausen1996emergence}. Therefore massive interests have been attracted to find a dictionary that can sparsely represent the training data. More specifically, dictionary learning is a bilinear inverse problem where both the dictionary and its corresponding sparse representations are to be learned. 

A typical dictionary learning algorithm is an iterative process alternating between two stages: sparse coding and dictionary update \cite{engan1999method,aharon2006k,engan2007family,skretting2010recursive,dai2012simultaneous,yu2019bilinear}. 
As dictionary learning involves two unknown variables, the general principle is to fix one variable and optimising the other. 
The purpose of sparse coding is to find the sparse coefficients based on a given dictionary. 
This optimisation problem can be solved using two different strategies: greedy algorithms such as matching pursuit (MP) \cite{mallat1993matching}, orthogonal matching
pursuit (OMP) \cite{pati1993orthogonal,tropp2007signal}, subspace
pursuit (SP) \cite{dai2009subspace}, CoSaMP \cite{needell2009cosamp} that select the support set from the sparse coefficients sequentially, and Basis Pursuit (BP) \cite{chen2001atomic} that convexifies the problem by replacing $\ell_0$ pseudo-norm with $\ell_1$ norm. The other stage dictionary update aims to refine the atoms of the dictionary using the sparse coefficients obtained from the previous stage. Method of optimal directions (MOD) \cite{engan1999method} is one of the earliest two staged methods, where the whole sparse coefficient matrix is fixed and the dictionary update is formulated as a least squares problem. In many other methods including K-SVD \cite{aharon2006k}, SimCO \cite{dai2012simultaneous}, and Blotless \cite{yu2019bilinear}, only the sparsity pattern (the positions of non-zeros) of sparse coefficients is fixed, and both the dictionary and the sparse coefficients are updated. Specifically, K-SVD updates one column of the dictionary and the corresponding row of sparse coefficients, while fixing all other dictionary atoms and the corresponding sparse coefficients. SimCO updates the whole dictionary and the whole sparse coefficient matrix  by viewing coefficients as a function of dictionary and performing a gradient descent with respect to dictionary. Blotless updates a block of the dictionary and the corresponding sparse coefficients using a total least squares approach.

\parskip = 0pt
In this paper, a novel dictionary learning algorithm that uses rank-one projection (ROP) is proposed. 
The key novelty in ROP is to formulate dictionary learning as an optimisation problem involving only one unknown variable, i.e., a set of rank-one matrices. 
That is, dictionary learning is cast as representing training data as the sum of  rank-one matrices, each with only a few non-zero columns. 
With this formulation, the two staged optimisation procedure in the literature is replaced with a single staged process. 
Then the popular alternating direction method of multipliers (ADMM) is adapted to solve the ROP formulation. 
Note that ROP involves a constrained optimisation with non-smooth objective function and a non-convex constraint (the set of rank-one matrices is non-convex). Nevertheless, recent advance in optimisation theory \cite{wang2019global} shows that the ADMM solver of ROP enjoys a global convergence guarantee. 

The single variable formulation ROP brings significant benefits. 
Firstly, it reduces the burden of parameter tuning. 
In the sparse coding stage of benchmark algorithms, one typically needs by trial-and-error to choose either the maximum sparsity level for greedy algorithms or a regularisation constant for a Lasso type of formulation. By comparison, there is no parameter to tune in ROP in generating all the simulations in this paper.
Secondly, our numerical results demonstrate that ROP outperforms other benchmark algorithms for the tests involving both synthetic data and real data. The results show that ROP can train good dictionaries with much less training data compared with other benchmark algorithms. In real data test, the performance improvement of ROP is demonstrated using examples of single image super-resolution. 

\section{Background \label{sec:background}}

\noindent
The goal of dictionary learning is to seek a dictionary that can sparsely represent the training data. Let $\bm{Y}\in\mathbb{R}^{M \times N}$, where $M\in\mathbb{N}$ and  $N\in\mathbb{N}$ denote the dimension and the number of training vectors, respectively. Dictionary learning can be written as
\begin{equation} \label{eq:dict-learning}
\underset{\bm{D},\bm{X}}{\min}\; \sum_n \left\Vert \bm{X}_{:,n}\right\Vert _{0}\;{\rm s.t.}\;\bm{Y} \approx \bm{D}\bm{X},
\end{equation} 
where $\bm{D}\in\mathbb{R}^{M\times K}$ denotes the unknown dictionary, and $\bm{X}\in \mathbb{R}^{K\times N}$ are the sparse representation coefficients, $\bm{X}_{:,k}$ is the $k$-th column of the matrix $\bm{X}$, and $\Vert \cdot \Vert_0$ is the $\ell_0$ pseudo-norm. The constraint $\bm{Y} \approx \bm{D}\bm{X}$ can be rewritten as $\Vert \bm{Y}-\bm{D}\bm{X}\Vert_F \le \epsilon$ when the noise energy in the training data can be roughly estimated, where $\Vert \cdot \Vert_F$ denotes the Frobenius norm and $\epsilon>0$ is a constant chosen based on the noise energy. In dictionary learning problems, it is typical that $M<K$, i.e., the dictionary is over-complete. 

The optimisation problem \eqref{eq:dict-learning} is non-convex due to the non-convexity of both the objective function and the constraint set. 
To make dictionary learning feasible, in the literature relaxation and/or extra constraints are imposed and suboptimal algorithms are designed \cite{engan1999method,aharon2006k,dai2012simultaneous,yu2019bilinear}. 
Note the scaling ambiguity that $\bm{D}_{:,k} \bm{X}_{k,:} = (a \bm{D}_{:,k}) (\frac{1}{a} \bm{X}_{k,:} )$, where $\bm{X}_{k,:}$ refers to the $k$-th row of the matrix $\bm{X}$. It is common to assume unit $\ell_2$-norm of columns of $\bm{D}$. Replacing the non-convex objective function in \eqref{eq:dict-learning} with the sparsity promoting $\ell_1$-norm, one has 
\begin{equation}
    \underset{\bm{D},\bm{X}}{\min} \sum_{n} \left\Vert\bm{X}_{:,n}\right\Vert_{1} \; {\rm s.t.} \; \bm{Y}\approx \bm{D}\bm{X},\;\Vert \bm{D}_{:,k} \Vert_2 = 1, \; \forall k\in [K], \label{eq:DL-l1}
\end{equation}
where $[K]:={1,2,\cdots,K}$. 

Another popular approach in the literature is to assume that the sparse representation of each training vector in $\bm{Y}$ has at most $S$ many non-zeros, where $S\in\mathbb{N}$ is a pre-defined constant typically carefully chosen by trial-and-error and usually $S \ll M$. The optimisation problem then becomes  
\begin{align}
\underset{\bm{D},\bm{X}}{\min}\; & \left\Vert \bm{Y}-\bm{D}\bm{X}\right\Vert _{2}^{2}\nonumber \\
{\rm s.t.}\; & \Vert\bm{D}_{:,k}\Vert_{2}=1,\;\Vert\bm{X}_{:,n}\Vert_{0}\le S,\;\forall n\in[N],\forall k\in[K].\label{eq:DL-sparsity-k}
\end{align}

Both problems \eqref{eq:DL-l1} and \eqref{eq:DL-sparsity-k} are typically solved by iterative algorithms that alternate between two stages: sparse coding and dictionary update. For simplicity, let us focus on solving Problem \eqref{eq:DL-sparsity-k} for now. In the sparse coding stage, one fixes the dictionary $\bm{D}$ and updates the coefficients $\bm{X}$ by 
\begin{equation}
\underset{\bm{X}_{:,n}}{\min}\;\Vert\bm{Y}_{:,n}-\bm{D}\bm{X}_{:,n}\Vert_{2}^{2},\;{\rm s.t.}\;\Vert\bm{X}_{:,n}\Vert_{0}\le S,\;\forall n\in [N]. \label{eq:sparse-coding}
\end{equation}
The non-convex problem \eqref{eq:sparse-coding} can be solved by many pursuit algorithms \cite{mallat1993matching,pati1993orthogonal,tropp2007signal,dai2009subspace,needell2009cosamp}. In the dictionary update stage, one updates the dictionary by fixing either the sparse coefficients, for example MOD \cite{engan1999method}, or the locations of the sparse coefficients, for example K-SVD \cite{aharon2006k}, SimCO \cite{dai2012simultaneous}, and Blotless \cite{yu2019bilinear}.

These two-staged algorithms have the same issue. The performance of the two stages are coupled together and the optimal tuning of one stage may not lead to the optimal performance of the overall dictionary learning. Furthermore, the two-stage alternating process makes the analysis very challenging. Few performance guarantees have been obtained in the literature for the general dictionary learning problem.

\section{Dictionary Learning Via ROP \label{sec:ROP-formulation}}

\noindent
This section derives the ROP formulation for dictionary learning which avoids alternating between two stages. 

We start with the constraint set in the original dictionary learning problem \eqref{eq:dict-learning}. It is straightforward to see that 
\begin{equation}
\bm{Y}\approx\bm{D}\bm{X}=\sum_{k}\bm{D}_{:,k}\bm{X}_{k,:}=\sum_{k}\bm{Z}_{k},
\end{equation}
where $ \bm{Z}_k := \bm{D}_{:,k}\bm{X}_{k,:}$ is a rank-one matrix for all $k\in [K]$. Define the set of rank-one matrices of proper size 
\begin{equation}
\mathcal{R}1=\left\{ \bm{Z}\in\mathbb{R}^{M\times N}:\;{\rm rank}\left(\bm{Z}\right)=1\right\} .
\end{equation}
Then the constraint set in \eqref{eq:dict-learning} can be written as 
\begin{equation}
\bm{Y} \approx \sum_{k}\bm{Z}_{k}, \; \bm{Z}_k\in \mathcal{R}1, \; \forall k\in[K]. \label{eq:ROP-constraint}
\end{equation}

The objective function is adapted accordingly. It is clear that a zero entry in $\bm{X}$, say $X_{k,n}$, results in a zero column in $\bm{Z}_k$, i.e., $(\bm{Z}_k)_{:,n} = \bm{D}_{:,k} X_{k,n} = \bm{0}$. The objective function is designed to promote zero columns in $\bm{Z}_k$, that is, $
\sum_{k} \left\Vert \bm{Z}_{k}\right\Vert_{2,0}$,
where 
\begin{equation*}
\left\Vert \bm{Z}_{k}\right\Vert _{2,0}:=\Vert[\Vert(\bm{Z}_{k})_{:,1}\Vert_{2}, \Vert(\bm{Z}_{k})_{:,2}\Vert_{2}\cdots,\Vert(\bm{Z}_{k})_{:,N}\Vert_{2}]^{T}\Vert_{0}
\end{equation*}
counts the number of non-zero columns of $\bm{Z}_k$. In practice, the non-convex $\ell_0$ pseudo-norm is replaced with convex $\ell_1$-norm, resulting in the convex objective function  
\begin{equation} \label{eq:ROP-objective}
\sum_{k}\left\Vert \bm{Z}_{k}\right\Vert _{2,1} := \sum_{k} \left( \sum_{n} \Vert(\bm{Z}_{k})_{:,n}\Vert_{2} \right). 
\end{equation}

Then dictionary learning is cast as
\begin{equation} \label{eq:ROP-formulation}
\min_{\bm{Z}_{k}}\;\sum_{k}\left\Vert \bm{Z}_{k}\right\Vert _{2,1}\;{\rm s.t.}\;\bm{Y}\approx\sum_{k}\bm{Z}_{k},\;\bm{Z}_{k}\in\mathcal{R}1,\;\forall k\in\left[K\right].
\end{equation}
The solutions of \eqref{eq:ROP-formulation} are invariant under the projection onto the set of rank-one matrices. Hence we term this formulation Rank-One Projection (ROP). It is a non-convex optimisation as the set $\mathcal{R}1$ is non-convex.
After solving \eqref{eq:ROP-formulation}, the dictionary items $\bm{D}_{:,k}$ and the corresponding coefficients $\bm{X}_{k,:}$ can be obtained using singular value decomposition (SVD) of $\bm{Z}_k$. 

Readers may wonder why not convexify the formulation \eqref{eq:ROP-formulation}. One way to convexity \eqref{eq:ROP-formulation} is to remove the constraint $\bm{Z}_{k}\in\mathcal{R}1$ and add a term $\mu \sum_k \Vert \bm{Z}_k \Vert_*$ into the objective function where $\mu>0$ is a regularisation constant and $\Vert \cdot \Vert_*$ denotes nuclear norm which promotes low rank structure. A careful analysis reveals that the above convexification will not lead to a dictionary that sparsely represents the training data in a proper way. The detailed analysis is omitted here due to the space constraint. 

It is worth to note that the ROP formulation \eqref{eq:ROP-formulation} avoids the scaling ambiguity in the formulations directly involving $\bm{D}$ and $\bm{X}$. Nevertheless like all other formulations, there is a permutation ambiguity in the solutions of \eqref{eq:ROP-formulation}. 

\subsection{An ADMM Solver for ROP \label{subsec:ADMM-solver}}

\noindent
In this subsection, ADMM technique is adapted to solve \eqref{eq:ROP-formulation}. As we show later, though the problem \eqref{eq:ROP-formulation} is non-convex, the ADMM procedure will converge. 

For compositional convenience, we derive ADMM version of ROP by replacing the constraint $\bm{Y}\approx \sum_k \bm{Z}_k$ in \eqref{eq:ROP-formulation} with $\bm{Y} = \sum_k \bm{Z}_k$. The extension to the constraint $\Vert \bm{Y} - \sum_k \bm{Z}_k \Vert_F \le \epsilon$ is similar and omitted here. Towards this end, we define the following indicator function for the set of rank-one matrices as 
\begin{equation}
\mathbbm{1}_{\mathcal{R}1}(\bm{Z})=\begin{cases}
\begin{array}{ll}
0,\\
+\infty,
\end{array} & \begin{array}{c}
{\rm if} \; \bm{Z}\in\mathcal{R}1,\\
\mathrm{otherwise}.
\end{array}\end{cases}
\end{equation}
Further, we introduce auxiliary variables $\bm{P}_k\in \mathbb{R}^{M\times N}$ and $\bm{Q}_K\in \mathbb{R}^{M\times N}$. An equivalent form to \eqref{eq:ROP-formulation} is given by
\begin{align}
& \min_{\bm{P}_{k},\bm{Q}_{k},\bm{Z}_{k}}\; \sum_{k} \Vert \bm{Q}_{k} \Vert _{2,1} + \sum_{k} \mathbbm{1}_{\mathcal{R}1}(\bm{Z}_{k}) \nonumber \\
& {\rm s.t.}\; \bm{Y}=\sum_{k}\bm{P}_{k},\;\bm{Q}_{k}=\bm{P}_{k},\;\bm{Z}_{k}=\bm{P}_{k},\:\forall k\in[K].\label{eq:ADMM-EasyForm}
\end{align}

For readability, we use the notations in \eqref{eq:ADMM-EasyForm} instead of standard ADMM form, and we derive detailed ADMM iteration steps as follow. As there are $MN+2MNK$ many equality constraints in \eqref{eq:ADMM-EasyForm}, we denote the corresponding Lagrange multipliers by $\bm{\Lambda}_0 \in \mathbb{R}^{M \times N}$, $\bm{\Lambda}_{1,k} \in \mathbb{R}^{M \times N}$, and $\bm{\Lambda}_{2,k} \in \mathbb{R}^{M \times N}$, referring to the equality constraints $\bm{Y}=\sum_k \bm{P}_k$, $\bm{Q}_k = \bm{P}_k$, and $\bm{Q}_k = \bm{Z}_k$, respectively. The ADMM iterations are given by 
\begin{align}
& (\cdots,\bm{P}_{k}^{l+1},\cdots) = \underset{\cdots,\bm{P}_{k},\cdots}{\arg\min} \; \Vert \sum_{k} \bm{P}_{k} -\bm{Y} +\bm{\Lambda}_{0}^{l} \Vert_{F}^{2} \nonumber \\
& + \sum_{k} \Vert \bm{P}_{k}-\bm{Q}_{k}^{l}+\bm{\Lambda}_{1,k}^{l} \Vert_{F}^{2} +\sum_{k} \Vert \bm{P}_{k}-\bm{Z}_{k}^{l}+\bm{\Lambda}_{2,k}^{l} \Vert_{F}^{2}, \label{eq:ADMM-P-update}
\end{align}
\begin{equation}
\bm{Q}_{k}^{l+1}=\underset{\bm{Q}_{k}}{\arg\min}\;\Vert\bm{Q}_{k}\Vert_{2,1}+\frac{\rho}{2}\Vert \bm{P}_{k}^{l+1}-\bm{Q}_{k}+\bm{\Lambda}_{1,k}^{l}\Vert _{F}^{2},\label{eq:ADMM-Q-update}
\end{equation}
\begin{equation}
\bm{Z}_{k}^{l+1} = \underset{\bm{Z}_{k}}{\arg\min} \;\mathbbm{1}_{\mathcal{R}1}(\bm{Z}_{k}) +\frac{\rho}{2} \Vert  \bm{P}_{k}^{l+1} -\bm{Z}_{k} + \bm{\Lambda}_{2,k}^{l} \Vert _{F}^{2},\label{eq:ADMM-Z-update}
\end{equation}
\begin{align}
\bm{\Lambda}_{0}^{l+1} & =\bm{\Lambda}_{0}^{l}+\sum_{k}\bm{P}_{k}^{l+1}-\bm{Y},\label{eq:ADMM-Lambda0-Update}\\
\bm{\Lambda}_{1,k}^{l+1} & =\bm{\Lambda}_{1,k}^{l}+\bm{P}_{k}^{l+1}-\bm{Q}_{k}^{l+1},\label{eq:ADMM-Lambda1-Update}\\
\bm{\Lambda}_{2,k}^{l+1} & =\bm{\Lambda}_{2,k}^{l}+\bm{P}_{k}^{l+1}-\bm{Z}_{k}^{l+1},\label{eq:ADMM-Lambda2-Update}
\end{align}
where $l$ denotes the iteration number.

Each iteration of ADMM involves three optimisation problems (\ref{eq:ADMM-P-update}-\ref{eq:ADMM-Z-update}) that are conceptually easy to solve. The optimisation problem \eqref{eq:ADMM-P-update} is a quadratic programming. In principle, it can be solved by many commercially available optimisation toolkits. However, this quadratic programming involves $MNK$ many unknowns and a linear map of dimension $(MN+2MNK)\times MNK$. Its huge dimension results in large run-time when using standard solvers. To address this problem, we design and implement a conjugate gradient (CG) procedure which uses the structures in \eqref{eq:ADMM-P-update} to simplify the computations substantially. In our simulation part, our CG procedure cuts the run-time in orders of magnitude. The details are omitted here due to the length constraint of this paper. 

The optimisation problem \eqref{eq:ADMM-Q-update} is convex but involves a non-differential term $\Vert \cdot \Vert_{2,1}$ in its objective function. The closed form of the optimal solution of \eqref{eq:ADMM-Q-update} can be obtained by setting the sub-gradient of the objective function to zero. Define $\hat{\bm{Q}}_k := \bm{P}_{k}^{l+1} + \bm{\Lambda}_{1,k}^{l}$. Then 
\begin{equation}
(\bm{Q}_{k}^{l+1})_{:,n}=\left(1-\frac{1}{\rho\Vert(\hat{\bm{Q}}_{k})_{:,n}\Vert_{2}}\right)_{+}(\hat{\bm{Q}}_{k})_{:,n},\label{eq:Q-solution}
\end{equation}
where $(x)_+ := \max(0,x)$. 

The optimisation problem \eqref{eq:ADMM-Z-update} is non-convex. Fortunately by Eckart-Young-Mirsky theorem, it can be solved by using singular value decomposition (SVD). Define $\hat{\bm{Z}}_k = \bm{P}_{k}^{l+1} +\bm{\Lambda}_{2,k}^{l}$. Consider the SVD of the matrix $\hat{\bm{Z}}_k$. Denote its largest singular value by $\sigma_1$ and the corresponding right and left singular vectors by $\bm{u}_1$ and $\bm{v}_1$ respectively. Then 
\begin{equation}
\bm{Z}_{k}^{l+1}=\sigma_{1}\bm{u}_{1}\bm{v}_{1}^{T}.\label{eq:Z-solution}
\end{equation}

\subsection{Convergence of ROP \label{subsec:Convergence}}

\noindent
ROP involves a non-convex ADMM with a non-smooth objective function. It is important to ensure its convergence before using it in practice. From the results in \cite[Theorem 1]{wang2019global}, our ROP algorithm indeed enjoys the global convergence guarantee. 

To apply the results in \cite{wang2019global}, the following definition is needed. 
\begin{definition}{(Restricted prox-regularity) \cite[Definition 2]{wang2019global}}
	For a lower semi-continuous function $f$, let $J\in\mathbb{R}_{+}$, $f:\mathbb{R}^{N}\rightarrow\mathbb{R}\cup\{\infty\}$, and define the excusion set
	\begin{equation}
	S_{J}:=\{x \in \operatorname{domain}(f):\|d\|>J \text { for all } d \in \partial f(x)\}
	\end{equation}
	$f$ is called \textit{restricted prox-regular} if, for any $J>0$ and bounded set $T \subseteq \operatorname{domain} f$, there exists $\gamma>0$  such that
	\begin{align}
	&   f(y)+\frac{\gamma}{2}\|x-y\|^{2} \geq f(x)+\langle d, y-x\rangle,\nonumber\\
	&   \forall x \in T \backslash S_{J}, y \in T, d \in \partial f(x),\|d\| \leq J.
	\end{align} 
\end{definition}

Specifically, consider the ADMM formulation of ROP in \eqref{eq:ADMM-EasyForm}, it can be verified that the first term in the objective function $\sum_{k} \Vert \bm{Q}_{k} \Vert _{2,1}$ is restricted prox-regular. The second term in the objective function is an indicator function of rank-one matrices, which is lower  semi-continuous. According to the conditions in \cite[Theorem 1]{wang2019global}, the ADMM process defined in Section \ref{subsec:ADMM-solver} converges to a stationary point. 

\begin{table*}
\caption{\label{tab:1}Comparison of single image super-resolution using different
dictionary learning methods, where both the figures of super-resolution
results and the errors between the estimated high-resolution digits
and the ground truth digits are shown in the table.}
\renewcommand{\arraystretch}{1.2}
\begin{centering}
\begin{tabular}{|C{2cm}| C{2cm}| C{1.4cm} C{1.4cm} C{1.4cm} C{1.4cm}|}
\hline

High resolution & Low resolution  & \multicolumn{4}{c|}{Methods}\\\cline{3-6}

ground truth & samples & ROP & BLOTLESS & K-SVD & MOD\\

\includegraphics[scale=0.1]{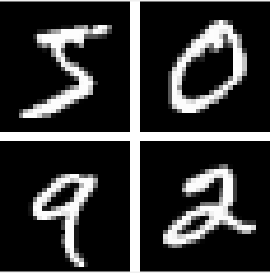}  &
\includegraphics[scale=0.1]{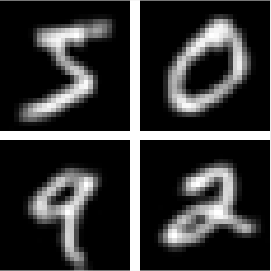}  &
\includegraphics[scale=0.1]{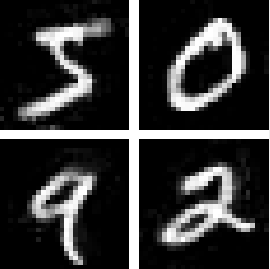} & \includegraphics[scale=0.1]{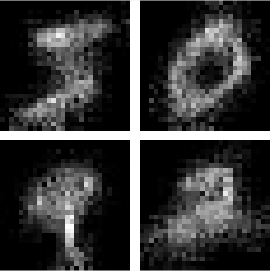} & \includegraphics[scale=0.1]{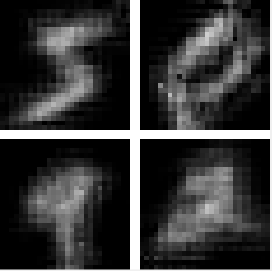} & 
\includegraphics[scale=0.1]{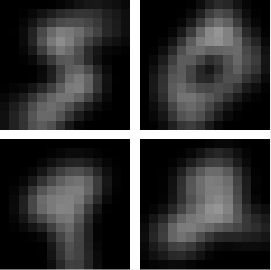}\\
\hline
Error of digit 5 & 0.1038 & 0.0464 & 0.3519 & 0.3264 & 0.4757\\
 
Error of digit 0 & 0.0900 & 0.0117 & 0.2919 & 0.4493 & 0.4626\\
 
Error of digit 9 & 0.1292 & 0.0471 & 0.3693 & 0.4229 & 0.4710\\
 
Error of digit 2 & 0.0977 & 0.0281 & 0.3528 & 0.3738 & 0.4302\\
\hline
\end{tabular}
\par\end{centering}
\end{table*}

\section{Numerical Tests}

\noindent
This section compares the numerical performance of ROP with other benchmark dictionary learning algorithms including MOD, K-SVD, and BLOTLESS. The comparison in Section \ref{subsec:synthetic-data} is based on synthetic data while real data is involved in Section \ref{subsec:real-data}. 

\subsection{Dictionary learning for synthetic data \label{subsec:synthetic-data}}

\noindent
For synthetic data tests, we adopt the typical setting for data generation. 
We assume that the training data $\bm{Y}$ are generated from a ground-truth dictionary $\bm{D}^0$ and a ground-truth sparse coefficient matrix $\bm{X}^0$ via $\bm{Y}=\bm{D}^0 \bm{X}^0$.
The dictionary $\bm{D}^0$ is generated by first filling it with independent realisations of the standard Gaussian variable and then normalising its columns to have unit $\ell_2$-norm. The sparse coefficients in $\bm{X}^0$ is generated as follows. Assume that the number of nonzero coefficients in the $n$-th column of $\bm{X}^0$ is $S_n$. The index set of the nonzero coefficients are randomly generated from the uniform distribution on $[K] \choose S_n$ and the values of the nonzero coefficients are independently generated from the standard Gaussian distribution. In our simulations, we set $S_{n} = S \in \mathbb{N}$, $\forall n\in [N]$.

Given the synthetic data, different dictionary learning algorithms are tested. OMP \cite{pati1993orthogonal} is used for the sparse coding stage of MOD, K-SVD, and BLOTLESS, with the prior knowledge of $S$. Note that different from other benchmark algorithms, ROP does not require such prior information. 

Dictionary learning algorithms are compared using dictionary recovery error. Consider the permutation ambiguity of the trained dictionary. The dictionary recovery error is defined as 
\begin{equation}
\mathrm{Error}\coloneqq\frac{1}{K}\sum_{k=1}^{K}(1-\mid\hat{\bm{D}}_{:,k}^{T}\bm{D}_{:,i_{k}}^{0}\mid),\label{eq:DRE}
\end{equation}
where $i_{k} \coloneqq{\arg\max}_{i\in\mathcal{I}_{k}}(\hat{\bm{D}}_{:,k}^{T}\bm{D}_{:,i}^{0})$, $\mathcal{I}_{k} \coloneqq[K]\backslash\{i_{1},\cdots,i_{k-1}\}$, $\hat{\bm{D}}_{:,k}$ denotes the $k$-th column of estimated
dictionary, and $\bm{D}_{:,i_{k}}^{0}$ represents the $i_{k}$-th
column of ground truth dictionary which has largest correlation with
$\hat{\bm{D}}_{k}$. The use of $\mathcal{I}_{k}$ is to avoid repetitions in $i_{k}$, $\forall k\in[K]$.

\begin{figure}
\begin{centering}
\subfloat[$M=16$, $K=32$, $S=3$.]{\begin{centering}
\includegraphics[scale=0.27]{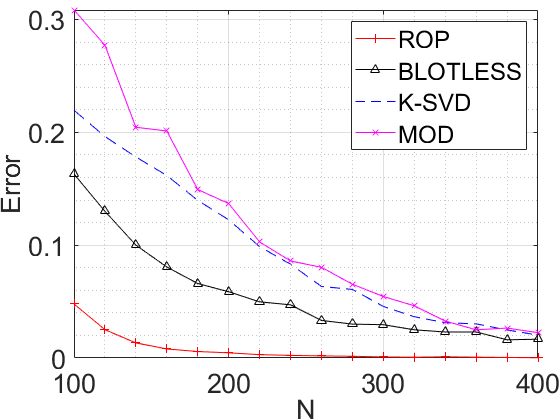}
\par\end{centering}
}\subfloat[$M=24$, $K=48$, $S=3$. ]{\begin{centering}
\includegraphics[scale=0.27]{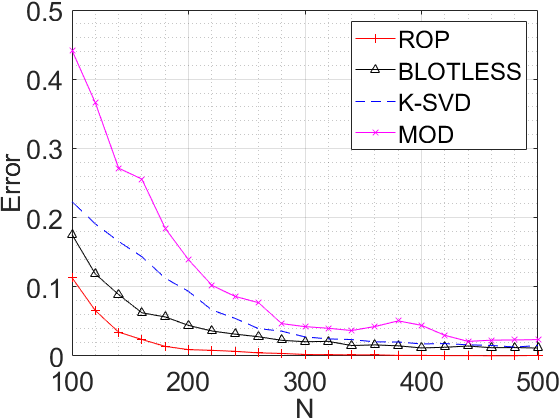}
\par\end{centering}
}
\par\end{centering}
\begin{centering}
\subfloat[$M=24$, $K=48$, $S=6$.]{\begin{centering}
\includegraphics[scale=0.27]{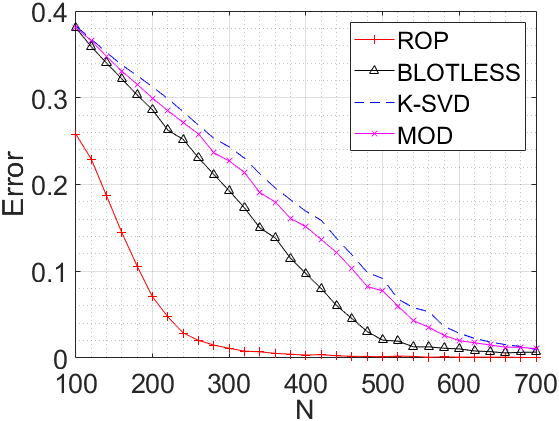}
\par\end{centering}
}\subfloat[$M=32$, $K=64$, $S=6$.]{\begin{centering}
\includegraphics[scale=0.27]{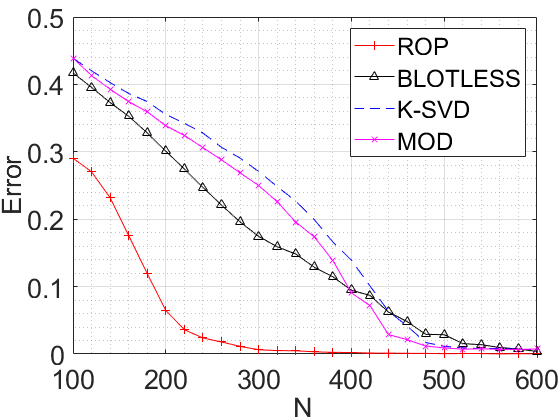}
\par\end{centering}
}
\par\end{centering}
\caption{\label{fig:1}Comparison of dictionary learning methods for the noise-free cases. Results are averages of 100 trials.}
\end{figure}

Fig. \ref{fig:1} compares the performance of dictionary learning algorithms. The  results are averages of 100 random trials, and in each trial the maximum number of iterations is set to 500. The results in Fig. \ref{fig:1} clearly show that ROP outperforms all other tested benchmark algorithms. The number of training samples required for ROP for a good recovery is the least. More importantly, when the number of training samples is relatively large, ROP is the only algorithm that has no visible error floor while all other algorithms suffer from non-negligible error floors.

\subsection{Single image super-resolution using dictionary learning \label{subsec:real-data}}

\noindent
This subsection focuses on the performance comparison of dictionary learning
algorithms when applied for single image super-resolution problem. We follow the approach by Yang et al. in \cite{yang2010image}. The basic idea is that given pairs of low and high resolution images as training data, a pair of dictionaries are learned so that sparse approximations of each pair of low/high resolution images share the same coefficients. For a test image of low resolution, one first finds its sparse representation under the low-resolution dictionary, and then apply the corresponding sparse coefficients to the high-resolution dictionary to generate a high resolution image. 

Our simulations are based on MNIST dataset which contains images for digits from 0 to 9. Each image is of $28\times 28$ pixels. We generate low-resolution images of size $14 \times 14$ by grouping adjacent $2\times 2$ pixels from original images and taking their average as one pixel. 

The training data used for dictionary learning is patch based. Patches of size $3\times 3$ are extracted from the low-resolution images with 2 pixel overlap in either direction for adjacent patches. Find the corresponding patches of size $6\times 6$ from the high-resolution images. Stack each pair of low and high resolution patches to form a column in the training data, i.e.,  $\bm{Y}_{:,n}=\left[{\rm vect}(\bm{P}_{L})_n^{T},{\rm vect}(\bm{P}_{H})_n^{T}\right]^{T}$, where $\bm{P}_L$ and $\bm{P}_H$ are low/high resolution patches respectively. In the simulations, we use 144 patches and hence the training sample matrix $\bm{Y}$ is of size $45 \times 144$. 

We then apply different algorithms for dictionary learning. Denote the acquired dictionary by $\bm{D}=\left[\bm{D}_{L}^{T},\bm{D}_{H}^{T}\right]^{T}$, where $\bm{D}_{L}$ and $\bm{D}_{H}$ are the sub-dictionaries corresponding to low and high resolution patches respectively. Here we set $K=128$. Given a low-resolution image for the test, extract $3\times 3$ patches with overlap of 2 pixels between adjacent patches in either direction. For each patch, a sparse representation coefficient vector $\bm{\alpha}$ is obtained so that $\bm{P}_L \approx \bm{D}_L \bm{\alpha}$  using sparse coding technique for example OMP. The corresponding high resolution patches are generated via $\bm{D}_H \bm{\alpha}$ and the high resolution image is generated by aligning the patches and taking average of overlapped pixels across patches. 

The simulation results are presented in Table \ref{tab:1}. In numerical comparison, normalised Frobenius norm $\Vert \hat{\bm{I}}-\bm{I}^{0} \Vert_{F}^{2}/\Vert \bm{I}^{0} \Vert_{F}^{2}$ is used as the performance criterion, where $\bm{I}^0$ and $\hat{\bm{I}}$ are the `ground-truth' high-resolution image and a high-resolution image generated using the learned dictionary, respectively. Simulation results demonstrate the significant improvement of ROP in both the numerical error and the visual effect.

\section{Conclusion}
\noindent
In this paper, we propose a novel dictionary learning algorithm using rank-one projection (ROP), where the problem is cast as an optimisation with respect to a single variable. Practically ROP reduces the number of tuning parameters required in other benchmark algorithms. An ADMM is derived to solve the optimisation problem and guarantees a global convergence. The test results show that ROP
outperforms other benchmark algorithms for both synthetic data and real data
especially when the number of sample is small.

\clearpage

\bibliographystyle{plain}
\bibliography{ICASSP2019ref}

\end{document}